\documentstyle[aps,preprint,eqsecnum,pra,epsf]{revtex}

\begin{document}

\draft

\tighten

\title{Spatial coherence and density correlations of trapped Bose gases}
\author{M.~Naraschewski}
\address{Jefferson Laboratory, Department of Physics, 
Harvard University, Cambridge MA~02138, USA}
\address{ITAMP, Harvard-Smithsonian Center for Astrophysics, 60 Garden Street,
Cambridge, MA~02138, USA}
\author{R.~J.~Glauber}
\address{Lyman Laboratory, Department of Physics, 
Harvard University, Cambridge MA~02138, USA}

\date{June 26, 1998}

\maketitle

\begin{abstract}
We study first and second order coherence of trapped dilute Bose gases using
appropriate correlation functions. Special attention is given to the discussion
of second order or density correlations.
Except for a small region around the surface of a Bose-Einstein
condensate the correlations can be accurately described as those of a locally
homogeneous gas with a spatially varying chemical potential. The degrees of
first and second order coherence
are therefore functions of temperature, chemical potential, and position.
The second order correlation function is governed both by the tendency of
bosonic atoms to cluster and by a strong 
repulsion at small distances due to atomic interactions. In present experiments
both effects are of comparable magnitude. Below the critical 
temperature the range of the bosonic correlation is affected by the presence 
of collective quasi-particle excitations. The results of some recent 
experiments on second and third order coherence are discussed. It is shown that
the relation between the measured quantities and the correlation functions 
is much weaker than previously assumed.
\end{abstract}


\narrowtext

\section{Introduction}
\label{Introduction}
The defining property of a Bose-Einstein condensate is the 
macroscopic occupation of a single one-particle quantum 
state \cite{penrose56}, a condition often referred to a off-diagonal long 
range order. 
Even though Bose-Einstein condensation has been long known to exist in systems
such as superfluid $^4$He, the off-diagonal long range order itself was
inaccessible to direct experimental verification. The  
achievement of Bose-Einstein condensation in dilute alkali gases 
\cite{anderson95,bradley95,davis95b}, however, has revived interest in such 
direct experimental access. Indeed, after a number of theoretical studies 
\cite{javanainen96a,naraschewski96a,hoston96,cirac96b,wong96a,wallis96,castin96b},
it has become possible to demonstrate  
off-diagonal long range order as made manifest by an interference
experiment \cite{andrews97} and that finding has been confirmed by 
excellent agreement between experiment and theory \cite{roehrl97a,wallis97b}. 

The ability of a Bose-Einstein condensate to produce an interference pattern
of high visibility is closely related to the first order coherence 
of an optical laser beam. More recently, two additional experiments have also
explored certain aspects of the second \cite{ketterle97} and third order
coherence \cite{burt97} of a trapped Bose condensed gas. 
It is therefore the aim of this paper to extend the theory of optical
coherence \cite{glauber63a} to the description of trapped dilute Bose gases.  
Our main focus will be on first and second order coherence.  

The preferred tools to study coherence are correlation functions. While 
first order coherence relies on the properties of the reduced single-particle
density matrix, which are expressed in the first order correlation function,
second order coherence depends mainly on density correlations.
Here, the strong short range repulsion between interacting atoms adds 
interesting new aspects to the discussion of atomic coherence, not present in 
its optical predecessor. Density correlations are an important object of study,
of course, even apart from their bearing on coherence. 

The paper is organized as follows. We introduce in Sec. \ref{Correlations}
appropriate first and second order correlation functions, which are then used
in Sec.\ \ref{Coherence} to define coherence. It is the aim of the remainder of
the paper to calculate these correlation functions for an externally trapped 
Bose gas. In most of our explicit calculations we assume a three-dimensional
harmonic trapping potential and evaluate our results in the thermodynamic 
limit, i.e., we will not account for
effects due to a finite atom number. This restriction is justified by the
large number of atoms $(N\approx 10^7)$ in recent experiments. 
We thereby show that coherence is not only a function of temperature and
chemical potential but also of the local position of the atoms.
We consider first in Sec.\ \ref{IdealGas} the case of an ideal Bose
gas. This allows us to study some basic aspects of the problem and in 
particular the applicability of a local density approximation.
Atomic interactions are added in Sec.\ \ref{InteractingGas}. 
The main modifications are brought about by the existence of quasi-particle 
like excitations and by the strong short range repulsion between the atoms. 
In Sec.\ \ref{CoherenceProp} we discuss the coherence properties that
follow from the correlation functions calculated earlier.
We also comment on 
the aforementioned measurements of higher order coherence 
\cite{ketterle97,burt97}
and a previous theoretical interpretation \cite{dodd97}. We summarize our 
conclusions in Sec.\ \ref{Conclusion}.

\section{Definitions}
\label{Definitions}

\subsection{Correlation functions}
\label{Correlations}
We describe the atoms by a quantum field 
$\hat\Psi({\bf r},t)$ which obeys the equal-time commutation relations
\begin{eqnarray}
[\hat\Psi({\bf r},t),\hat\Psi^\dagger({\bf r}^\prime,t)]&=&
\delta({\bf r}-{\bf r}^\prime)\\
\lbrack\hat\Psi({\bf r},t),\hat\Psi({\bf r}^\prime,t)\rbrack&=&0.
\end{eqnarray}

In general, physical properties of the gas can be expressed in 
terms of correlation functions, i.e., expectation values of the field 
operators of the type
\begin{equation}
\langle\hat\Psi^\dagger 
({\bf r}_1,t_1)\hat\Psi^\dagger({\bf r}_2,t_2)\cdots
\hat\Psi({\bf r}^\prime_2,t_2^\prime)
\hat\Psi({\bf r}^\prime_1,t_1^\prime)\rangle.
\end{equation}
Here, a description of the dynamics using the Heisenberg picture has been 
tacitly assumed. However, we shall restrict ourselves to discussing  
spatial correlations at thermal equilibrium and at a fixed time. We
therefore omit further mention of the time arguments. Temporal correlations
have been discussed recently in the context of condensate phase fluctuations
\cite{wallis96,castin96b,sols94,wright96,lewenstein96c,naraschewski96b,naraschewski97b,imamoglu97,jaksch97b}. 

For later convenience we introduce a special notation for the correlation 
functions 
\begin{eqnarray}
G^{(1)}({\bf r},{\bf r}^\prime)&=&\langle\hat\Psi^\dagger 
({\bf r})\hat\Psi({\bf r}^\prime)\rangle\\
G^{(2)}({\bf r},{\bf r}^\prime)&=&\langle\hat\Psi^\dagger 
({\bf r})\hat\Psi^\dagger({\bf r}^\prime)\hat
\Psi({\bf r}^\prime)\hat\Psi({\bf r})\rangle.
\end{eqnarray}
The first order correlation function $G^{(1)}({\bf r},{\bf r}^\prime)$ contains
the same information as the reduced one-particle density matrix which furnishes
the expectation values of
one-particle observables like the momentum or position density. The formal
relation between first order correlation function and reduced one-particle 
density matrix is given by $G^{(1)}({\bf r},{\bf r}^\prime)=
\rho^{(1)}({\bf r}^\prime,{\bf r})=\langle{\bf r}^\prime|\rho^{(1)}|{\bf r}
\rangle$.  
The first order correlation function can alternatively be expressed in terms 
of the Wigner function
\begin{equation}
W({\bf p},{\bf q})=\frac{1}{(2\pi\hbar)^3}\int\!d{\bf r}\, e^{-i{\bf p}\cdot
{\bf r}/\hbar}\,\langle\hat\Psi^\dagger 
({\bf q}-\frac{\bf r}{2})\hat\Psi({\bf q}+\frac{\bf r}{2})\rangle,
\end{equation}
which is a quantum mechanical analog of a classical single-particle 
phase space distribution.
The connection between the quantum mechanical correlation function and 
statistical mechanics is indeed illustrated by the Fourier transform relation
\begin{equation}
\label{G1Wigner}
G^{(1)}({\bf r},{\bf r}^\prime)=\int\!d{\bf p}\, e^{-i{\bf p}\cdot
({\bf r}-{\bf r}^\prime)/\hbar}\,W({\bf p},\frac{{\bf r}+{\bf r}^\prime}{2}). 
\end{equation}

In contrast, the second order correlation function 
$G^{(2)}({\bf r},{\bf r}^\prime)$ expresses the joint probability of detecting
two different atoms at the locations $\bf r$ and ${\bf r}^\prime$ respectively.
Its connection with the density correlation function is given by  
\begin{equation}
\label{ncorr}
\langle\hat\Psi^\dagger 
({\bf r})\hat\Psi({\bf r})\hat\Psi^\dagger({\bf r}^\prime)\hat
\Psi({\bf r}^\prime)\rangle 
= G^{(2)}({\bf r},{\bf r}^\prime)+G^{(1)}({\bf r},{\bf r})
\,\delta({\bf r}-{\bf r}^\prime).
\end{equation}
It is because the function $G^{(2)}$ is defined as the average of a normally 
ordered product of field operators, that it expresses the correlation of 
distinct pairs of atoms.
The delta function contribution in Eq.\ (\ref{ncorr}) represents by contrast
the autocorrelation of individual atoms. Discussions of the coherence 
properties of fields are most conveniently phrased in terms of factorization
of their correlation functions \cite{glauber63a}. The 
autocorrelation 
contribution in Eq.\ (\ref{ncorr}) however stands in the way of any rigorous
factorization of the density correlation into a product of functions of
$\bf r$ and ${\bf r}^\prime$. We shall therefore concentrate on discussing
the normally ordered correlation function $G^{(2)}({\bf r},{\bf r}^\prime)$
which can indeed meet the factorization condition under suitable physical
circumstances.

\subsection{Coherence and Bose-Einstein condensation}
\label{Coherence}
In the optical context coherence refers to a state of the system 
with interference properties that are as close as possible to those of a 
precisely defined classical field.
In more technical terms, successive orders of coherence can be described by
means of factorization conditions imposed upon a succession of field 
correlation functions. The analog of $n$-th order coherence for the atomic
field would require all normally ordered correlation functions of order
$m\le n$ to factorize into a form
\begin{equation}
\label{qocoherence}
\langle\hat\Psi^\dagger 
({\bf r}_1)\hat\Psi^\dagger({\bf r}_2)\cdots
\hat\Psi({\bf r}^\prime_2)\hat\Psi({\bf r}^\prime_1)\rangle=
\psi^*({\bf r}_1)\psi^*({\bf r}_2)\cdots\psi({\bf r}^\prime_2)
\psi({\bf r}^\prime_1),
\end{equation}
i.e., into products of the same complex valued function $\psi({\bf r})$.
If the field were in a pure coherent state, the factors $\psi({\bf r})$ 
could be regarded as field-expectation values $\langle\hat\Psi({\bf 
r})\rangle$. This identification, however, is by no means necessary to 
secure the factorizations of Eq.\ (\ref{qocoherence}). It would even
contradict the strict conservation of the total atom number.

To identify the function $\psi({\bf r})$ we begin by considering first order
coherence, which secures the maximum effects of intensity interference between
spatially separated parts of the system.  
In this sense, perfect first order coherence 
\begin{equation}
\label{O1coherence}
G^{(1)}({\bf r},{\bf r}^\prime)=\psi^*({\bf r})\psi({\bf r}^\prime)
\end{equation}
with
\begin{equation}
\psi({\bf r})=\sqrt{N}\,\langle{\bf r}|\phi\rangle
\end{equation}
will hold if all atoms occupy the same one-particle state $|\phi\rangle$.
Here $N$ denotes the total atom number of the system.
This definition is an idealization of the 
Penrose and Onsager definition of Bose-Einstein condensation \cite{penrose56} 
which requires the reduced one-particle density matrix to have a single
macroscopic eigenvalue. 
The function $\psi({\bf r})$ can thus be 
introduced without breaking the gauge symmetry associated with 
the conservation of total atom number. The modulus
of $\psi({\bf r})$ can be reexpressed in terms of the first order correlation
function as
\begin{equation}
|\psi({\bf r})|=\sqrt{G^{(1)}({\bf r},{\bf r})}.
\end{equation}
In addition to the factorization of $G^{(1)}$, perfect second order coherence 
implies the factorization condition
\begin{equation}
\label{O2coherence}
G^{(2)}({\bf r},{\bf r}^\prime)=|\psi({\bf r})|^2\,|\psi({\bf r}^\prime)|^2
\end{equation}
so that
\begin{equation}
G^{(2)}({\bf r},{\bf r}^\prime)
=G^{(1)}({\bf r},{\bf r})\,G^{(1)}({\bf r}^\prime,{\bf r}^\prime).
\end{equation}
First order coherence does not, of course, require second order coherence to
hold. It does, however, imply a weaker factorization property of the function
$G^{(2)}$ \cite{titulaer65} which includes the statement
\begin{equation}
G^{(2)}({\bf r},{\bf r}^\prime)=g_2\,
|\psi({\bf r})|^2\,|\psi({\bf r}^\prime)|^2,
\end{equation}
where $g_2$ is a non-negative real number. For example, a first order coherent
system with definite total atom number can achieve second order coherence,
i.e.\ $g_2=1$, only to an accuracy of order $O(1/N)$.

In fact, any real atomic system can exhibit coherence only in an approximate 
way. In order to define a local measure of coherence we introduce 
correlation functions which are normalized to attain unit modulus in the case
of perfect coherence. The degree of first order coherence, in this sense, is 
expressed by 
\begin{equation}
\label{g1localdef}
g^{(1)}({\bf r},{\bf r}^\prime)=\frac{G^{(1)}({\bf r},{\bf r}^\prime)}
{\sqrt{G^{(1)}({\bf r},{\bf r})}\sqrt{G^{(1)}({\bf r}^\prime,{\bf r}^\prime)}},
\end{equation}
while 
\begin{equation}
\label{g2localdef}
g^{(2)}({\bf r},{\bf r}^\prime)=\frac{G^{(2)}({\bf r},{\bf r}^\prime)}
{G^{(1)}({\bf r},{\bf r})\,G^{(1)}({\bf r}^\prime,{\bf r}^\prime)}
\end{equation}
is a certain measure of second order coherence.
In the language of a classical field theory $g^{(1)}$ characterizes local
fluctuations of the phase of the complex field amplitude, while $g^{(2)}$
is related to fluctuations of its modulus. The most important property
of $g^{(1)}$ is its relation to the contrast achievable in an
interference experiment \cite{hecht98}. The function $g^{(2)}$, on the other 
hand, expresses the tendency of atoms either to cluster or to remain
spatially separated.

In many experimental contexts, it will be difficult, to collect enough data 
to determine the dependence of the correlation functions on the two 
coordinates
$\bf r$ and ${\bf r}^\prime$. It may then be simpler to measure volume 
integrated correlation functions 
\begin{eqnarray}
\label{G1volume}
G^{(1)}({\bf r})&=&\int\!d{\bf R}\,\,G^{(1)}({\bf R}-\frac{\bf r}{2},
{\bf R}+\frac{\bf r}{2})\\
\label{G2volume}
G^{(2)}({\bf r})&=&\int\!d{\bf R}\,\,G^{(2)}({\bf R}-\frac{\bf r}{2},
{\bf R}+\frac{\bf r}{2}),
\end{eqnarray}
in which a spatial integration over the mean
coordinate has been performed. In condensed matter physics, for example,  
scattering experiments are often used to obtain information about the volume 
integrated second order correlation function. We define normalized versions
of these functions 
\begin{eqnarray}
\label{g1volume}
g^{(1)}({\bf r})&=&\frac{\int\!d{\bf R}\,\,G^{(1)}({\bf R}-\frac{\bf r}{2},
{\bf R}+\frac{\bf r}{2})}{\int\!d{\bf R}\,\sqrt{G^{(1)}({\bf R}-\frac{\bf r}{2}
,{\bf R}-\frac{\bf r}{2})}\,\sqrt{G^{(1)}({\bf R}+\frac{\bf r}{2},
{\bf R}+\frac{\bf r}{2})}}\\
\label{g2volume}
g^{(2)}({\bf r})&=&\frac{\int\!d{\bf R}\,\,G^{(2)}({\bf R}-\frac{\bf r}{2},
{\bf R}+\frac{\bf r}{2})}{\int\!d{\bf R}\,G^{(1)}({\bf R}-\frac{\bf r}{2},
{\bf R}-\frac{\bf r}{2})\,G^{(1)}({\bf R}+\frac{\bf r}{2},
{\bf R}+\frac{\bf r}{2})},
\end{eqnarray}
which can also be interpreted as weighted volume averages of the normalized 
correlation functions of Eqs.\ (\ref{g1localdef}) and (\ref{g2localdef}).

It will be the aim of the following sections to calculate the 
correlation functions we have defined for trapped dilute Bose gases at finite
temperatures. The main emphasis will be put on systems that are large enough
to allow a local density approximation.

\section{Ideal gases}
\label{IdealGas}
The basic concepts for calculating the correlation functions of a trapped
gas are most easily described for ideal gases with temperatures above the 
critical temperature. 
The presence of a Bose-Einstein condensate at temperatures below the 
critical temperature gives rise to additional features which will be
discussed in Sec.\ \ref{BelowTc}. The influence of atomic interactions on both
uncondensed and condensed gases will be later dealt with in Sec.\ 
\ref{InteractingGas}.

\subsection{Temperatures above $T_c$}
\label{AboveTc}
In the simplest case, we consider an ideal Bose gas at temperature $T$ 
trapped by an external potential $V({\bf r})$. In the grand canonical ensemble 
different one-particle energy eigenstates $u_{\bf n}({\bf r})$ of the trapping
potential are populated independently. As a result we can write
\begin{eqnarray}
\label{G1ideal}
G^{(1)}({\bf r},{\bf r}^\prime)&=&\sum_{\bf m} u^*_{\bf m}({\bf r})\,
u_{\bf m}({\bf r}^\prime)\,
\langle\hat a_{\bf m}^\dagger a_{\bf m}\rangle\\
G^{(2)}({\bf r},{\bf r}^\prime)&=&\sum_{\bf klmn} u^*_{\bf k}({\bf r})\,
u^*_{\bf l}({\bf r}^\prime)\,u_{\bf m}({\bf r}^\prime)\,u_{\bf n}({\bf r})\,
\langle\hat a_{\bf k}^\dagger\hat a_{\bf l}^\dagger\hat a_{\bf m}\hat 
a_{\bf n}\rangle\nonumber\\
\label{G2ideal}
&=&\langle\hat\Psi^\dagger({\bf r})\hat\Psi
({\bf r})\rangle\langle\hat\Psi^\dagger({\bf r}^\prime)\hat\Psi
({\bf r}^\prime)\rangle+ |\langle\hat\Psi^\dagger({\bf r}^\prime)\hat\Psi
({\bf r})\rangle|^2\\
&&+\sum_{\bf m}|u_{\bf m}({\bf r})|^2|u_{\bf m}({\bf r}^\prime)|^2\left(
\langle\hat a_{\bf m}^\dagger\hat a_{\bf m}^\dagger\hat a_{\bf m}\hat 
a_{\bf m}\rangle-2
\langle\hat a_{\bf m}^\dagger\hat a_{\bf m}\rangle\langle\hat a_{\bf m}^\dagger
\hat a_{\bf m}\rangle\right).\nonumber
\end{eqnarray}
The statistical independence we have assumed for the mode occupations would
not, on the other hand, 
hold precisely for a canonical ensemble, i.e., a system
with a fixed total atom number. However, the error in the 
correlation function induced thereby is only of the order $O(1/N)$ and can be 
neglected for large systems.

The occupation fluctuations characteristic of the grand canonical ensemble
make the last line of Eq.\ (\ref{G2ideal}) vanish. The second order 
correlation functions can then be written as
\begin{eqnarray}
\label{G2aboveTc}
G^{(2)}({\bf r},{\bf r}^\prime)&=& G^{(1)}({\bf r},{\bf r})\,
G^{(1)}({\bf r}^\prime,{\bf r}^\prime)
+ |G^{(1)}({\bf r},{\bf r}^\prime)|^2\\
\label{g2aboveTc}
g^{(2)}({\bf r},{\bf r}^\prime)&=& 1 + |g^{(1)}({\bf r},{\bf r}^\prime)|^2.
\end{eqnarray}
The second term of $G^{(2)}$ or $g^{(2)}$ is a quantum statistical exchange 
term. It manifests
a tendency of bosonic atoms to cluster together. In the case of fermions
this term would have a negative sign reflecting the influence of the Pauli 
exclusion principle.

Since we may freely use the grand canonical ensemble at temperatures above
$T_c$, we need only determine the first order correlation 
functions $G^{(1)}({\bf r},{\bf r}^\prime)$ and 
$g^{(1)}({\bf r},{\bf r}^\prime)$. Furthermore, no condensate exists in this
regime and the normalized correlation functions have the limiting values
\begin{eqnarray}
\label{g1limits}
g^{(1)}({\bf r},{\bf r})=1\qquad\lim_{|{\bf r}-{\bf r}^\prime|\to\infty}
g^{(1)}({\bf r},{\bf r}^\prime)=0\\
\label{g2limits}
g^{(2)}({\bf r},{\bf r})=2\qquad\lim_{|{\bf r}-{\bf r}^\prime|\to\infty}
g^{(2)}({\bf r},{\bf r}^\prime)=1.
\end{eqnarray}
In the following we compute the length scale of 
${\bf r}-{\bf r}^\prime$ over which the correlation functions decrease.

\subsubsection{Three-dimensional harmonic oscillator potential}
As a first step we calculate the first order correlation function for 
the case of a spherical harmonic oscillator potential $V({\bf r})=m\omega^2
r^2/2$ with angular trap frequency $\omega$. 
We therefore occupy the harmonic oscillator states in Eq.\ (\ref{G1ideal}) 
with a Bose-Einstein distribution, so that 
\begin{equation}
\label{G10BE}
G^{(1)}({\bf r},{\bf r}^\prime)=\sum_{\bf m} u_{\bf m}^*({\bf r})
u_{\bf m}({\bf r}^\prime)
\,\frac{z\,e^{-\beta\tilde\epsilon_{\bf m}}}
{1- z\,e^{-\beta\tilde\epsilon_{\bf m}}}
\end{equation}
with the inverse temperature $\beta=1/k_BT$, the fugacity $z=e^{\beta\mu}$,
and the chemical potential $\mu$.
The energies $\tilde\epsilon_{\bf m}$ have the zero point energy, 
$3\hbar\omega/2$, removed in order to limit the fugacity to values between
zero and one, as in a homogeneous gas. If the summand of Eq.\ (\ref{G10BE})
is expanded in powers of $z$, the remaining sum over the state indices $\bf m$
can be compactly expressed in terms of the single-particle propagator
$G_0({\bf r}^\prime, {\bf r},t) = \langle{\bf r}^\prime|
\exp(-iHt)|{\bf r}\rangle$ for motion governed by the harmonic oscillator 
Hamiltonian, $H$. We obtain 
\begin{eqnarray}
G^{(1)}({\bf r},{\bf r}^\prime)
&=&\sum_{\bf m} u_{\bf m}^*({\bf r})u_{\bf m}({\bf r}^\prime)
\,z\,e^{-\beta\tilde\epsilon_{\bf n}}\sum_{k=0}^\infty z^k
e^{-k\beta\tilde\epsilon_{\bf n}}\nonumber\\
\label{G1resumm}
&=&\sum_{k=1}^\infty z^k\,\exp(3 k\beta\hbar\omega/2)
\,G_0({\bf r}^\prime, {\bf r},t=-i\hbar k\beta)\\
&=&\sum_{k=1}^\infty z^k\,
\,\tilde G_0({\bf r}^\prime, {\bf r},k\beta\hbar\omega)\nonumber
\end{eqnarray}
where
\begin{equation}
\label{G0im}
\tilde G_0({\bf r}^\prime, {\bf r},\tau)=\left(
\frac{m\omega}{\pi\hbar(1-e^{-2\tau})}\right)^{3/2}\,\exp
\left[-\frac{m\omega}{\hbar}\frac{(r^2+ r^{\prime 2})(\cosh\tau -1)+
({\bf r}-{\bf r}^\prime)^2}{2\sinh\tau}\right]
\end{equation}
is derived from the Green's function for the three-dimensional harmonic
oscillator \cite{sakurai94}.
The fugacity $z$ must be determined self-consistently by using the relation
\cite{bagnato87}
\begin{equation}
N=\sum_{k=1}^\infty\frac{z^k}{(1-e^{-k\beta\hbar\omega})^3}
\end{equation}
in order to insure the proper total atom number $N$. 

A considerable simplification can be achieved
if the energy spacing of the trap levels is much smaller than the thermal 
energy,
i.e., $\beta\hbar\omega\ll 1$. In that case $\tilde G_0$ reduces to 
\begin{equation}
\tilde G_0({\bf r}^\prime, {\bf r},\tau)=\left(
\frac{m\omega}{2\pi\hbar\tau}\right)^{3/2}\,\exp
\left[-\frac{m\omega}{\hbar}\frac{(r^2+ r^{\prime 2})\tau/2+
({\bf r}-{\bf r}^\prime)^2/\tau}{2}\right]
\end{equation}
and the sum for $G^{(1)}$ can be written in the form
\begin{eqnarray}
G^{(1)}({\bf r},{\bf r}^\prime)&=&\frac{1}{\lambda_T^3}\sum_{k=1}^\infty
\frac{z^k}{k^{3/2}}\,\exp\left[-\frac{[V({\bf r})+V({\bf r}^\prime)]/2}{k_BT}
\right]^k
\,\exp\left[-\pi\frac{({\bf r}-{\bf r}^\prime)^2}{\lambda_T^2}\right]^{1/k}
\nonumber\\
\label{G1ho}
&=&\frac{1}{\lambda_T^3}g_{3/2}\left(\exp\left[\frac{\mu-[V({\bf r})+V({\bf r}
^\prime)]/2}{k_BT}\right],\exp\left[-\pi\frac{({\bf r}-{\bf r}^\prime)^2}
{\lambda_T^2}\right]\right)
\end{eqnarray}
where we have used the thermal wavelength
\begin{equation}
\lambda_T=\hbar\sqrt{2\pi/mk_BT}.
\end{equation}
In addition we have introduced the generalized Bose function
\begin{equation}
g_\alpha(x,y)=\sum_{k=1}^\infty\frac{x^ky^{1/k}}{k^\alpha}
\end{equation}
which is related to the commonly used Bose function $g_{\alpha}(x) =
\sum_{k=1}^\infty x^k/k^\alpha$ and to the Riemann zeta function 
$\zeta(\alpha)=\sum_{k=1}^\infty 1/k^\alpha$ through
\begin{equation}
g_{\alpha}(x)=g_{\alpha}(x,1),\qquad \zeta(\alpha)=g_{\alpha}(1)=g_{\alpha}
(1,1).
\end{equation}

\subsubsection{Local density approximation}
The implications of Eq.\ (\ref{G1ho}) can be made clearer by deriving an
approximate form for it from Eq.\ (\ref{G1Wigner}). There $G^{(1)}$ was 
expressed in terms
of the Wigner function. We now employ the local density approximation which
assumes the Wigner function, $W({\bf p},{\bf q})$, to be locally identical to 
the momentum distribution of a spatially homogeneous system with constant 
potential energy equal to the local value $V({\bf q})$
\begin{equation}
W({\bf p},{\bf q})=\frac{1}{(2\pi\hbar)^3}\frac{1}{\exp\left[
(p^2/2m+V({\bf q})-\mu)/k_BT\right]-1}.  
\end{equation}
This approximation implies that the atoms see an effective chemical 
potential $\mu({\bf q})=\mu-V({\bf q})$ which varies with position.
If we Fourier transform this Wigner function as required by  
Eq.\ (\ref{G1Wigner}) the correlation function can be expressed as
\begin{equation}
\label{G1lda}
G^{(1)}({\bf r},{\bf r}^\prime)=\frac{1}{\lambda_T^3}g_{3/2}\left(\exp\left[
\frac{\mu-V[({\bf r}+{\bf r}^\prime)/2]}{k_BT}\right],\exp\left[-\pi
\frac{({\bf r}-{\bf r}^\prime)^2}{\lambda_T^2}\right]\right).
\end{equation}
The global chemical potential $\mu$ must then be chosen to secure the
correct total atom number 
\begin{equation}
\label{Nsum}
N=\int\!d{\bf r}\,G^{(1)}({\bf r},{\bf r}).
\end{equation}

We can now compare Eq.\ (\ref{G1lda})
with Eq.\ (\ref{G1ho}). These expressions for $G^{(1)}$ in fact coincide
for equal arguments ${\bf r}={\bf r}^\prime$. 
However, for ${\bf r}\neq{\bf r}^\prime$ they treat
the potential energy in a slightly different way. The accuracy of the 
local density approximation evidently requires that the external potential 
vary slowly on a length scale given by $\lambda_T$. For a
relative separation ${\bf r}-{\bf r}^\prime$ of this magnitude,
the difference of the potential energy between Eqs.\ (\ref{G1ho}) and 
(\ref{G1lda}) is only of the order $(\hbar\omega)^2/k_BT$ and can thus be
neglected. This result can be no surprise if we consider that the 
thermal wavelength, written as $\lambda_T=\sqrt{2\pi\hbar\omega/k_BT}\,a_0$, 
becomes smaller than the natural length unit 
$a_0=\sqrt{\hbar/m\omega}$ of the trapping potential precisely
in the limit $\hbar\omega\ll k_BT$. For the trap 
parameters of Ref.\ \cite{andrews97}, the thermal wavelength assumes a value
of $\lambda_T=0.4\,\mu$m at the critical temperature, which is three to ten 
times smaller than the harmonic oscillator length units of the anisotropic 
trapping potential.

We have thus shown that both the density distribution and 
correlations implicit in $G^{(1)}$ are well described by the local density 
approximation, as long as
the separation of the trap levels is small compared to the thermal energy.
Of course, application of the local density approximation
is not restricted to an isotropic harmonic oscillator potential. It can be
applied to any trapping potential $V({\bf r})$ as long as the thermal 
wavelength is
small compared to the length scale of the spatial variations of the potential.
Since the correlation function decreases on the scale of the thermal 
wavelength it is consistent to use for small relative separations ($r<a_0$)
the alternative definitions
\begin{eqnarray}
\label{g1volumealt}
g^{(1)}({\bf r})&=&\frac{\int\!d{\bf R}\,\,G^{(1)}({\bf R}-\frac{\bf r}{2},
{\bf R}+\frac{\bf r}{2})}{\int\!d{\bf R}\,G^{(1)}({\bf R}
,{\bf R})}\\
\label{g2volumealt}
g^{(2)}({\bf r})&=&\frac{\int\!d{\bf R}\,\,G^{(2)}({\bf R}-\frac{\bf r}{2},
{\bf R}+\frac{\bf r}{2})}{\int\!d{\bf R}\,|G^{(1)}({\bf R},{\bf R})|^2}
\end{eqnarray}
for the normalization of the volume integrated correlation functions.

\begin{figure}
\begin{center}
\leavevmode
\epsfxsize=0.5\textwidth
\epsffile{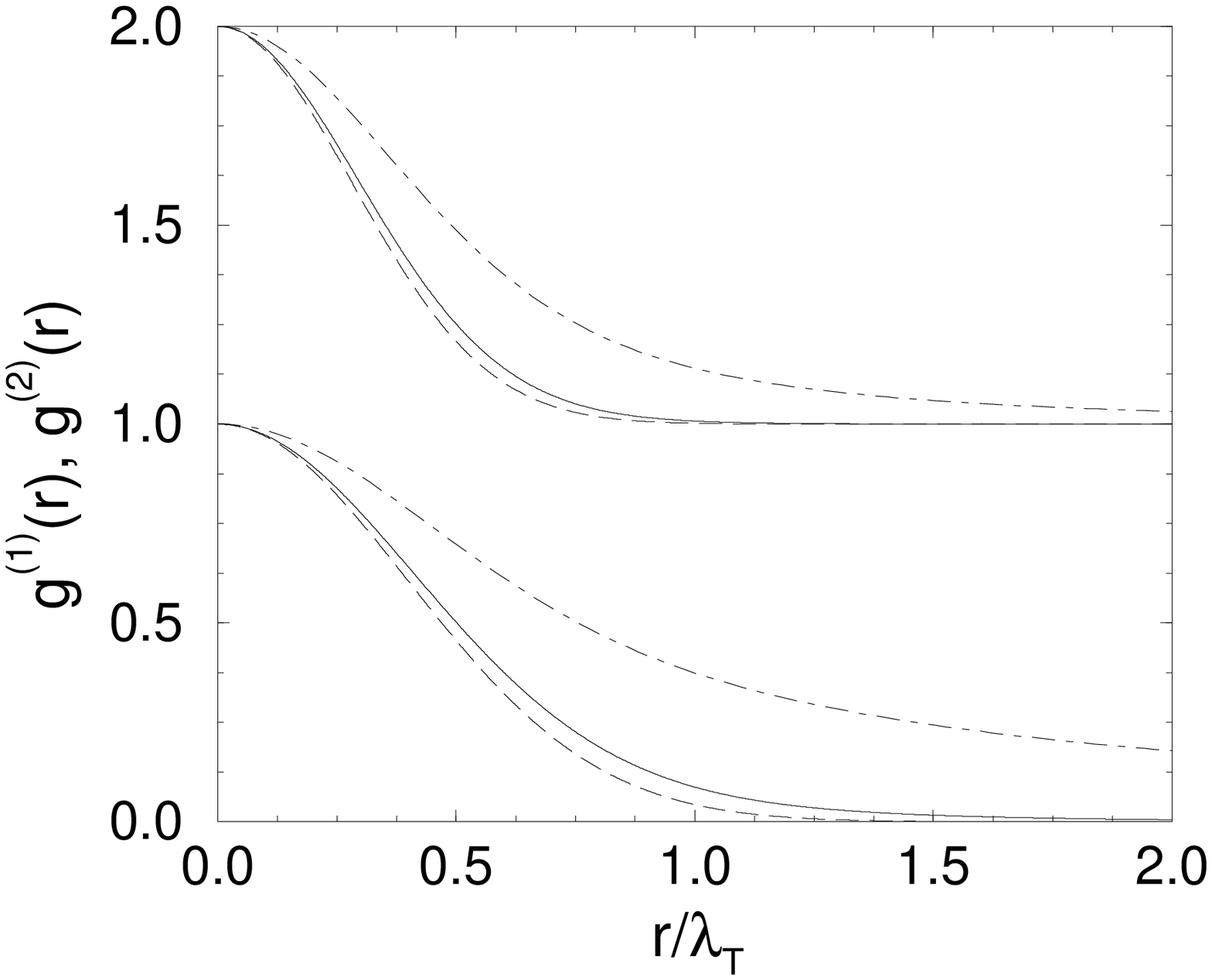}
\end{center}
\caption{\label{gidealfig}This plot shows the volume integrated first order 
(bottom)
and second order (top) correlation functions for a harmonically trapped ideal
gas at the critical temperature. The solid lines correspond to the exact
expressions of Eqs.\ (\ref{g1volumeharm}) and (\ref{g2aboveTc}), while the 
dashed lines approximate the occupation distribution of the trap levels by a 
Maxwell-Boltzmann distribution as in Eq.\ (\ref{gmb}). In order to display
the diluting effect of the volume integration, implicit in these results, 
we have added as dot-dashed lines the unintegrated 
correlation functions for mean positions at the center of the trap, where the 
local chemical potential is the smallest.}  
\end{figure}

For a harmonic oscillator potential the integration over the mean coordinate 
${\bf R}=({\bf r}+{\bf r}^\prime)/2$ of
Eq.\ (\ref{G1lda}) can easily be carried out \cite{giorgini97}
by using the series expansion of $g_{3/2}$. The results for  
the volume integrated correlation functions of Eqs.\ (\ref{G1volume}) and 
(\ref{g1volumealt}) are then 
\begin{eqnarray}
G^{(1)}({\bf r})&=&\left(\frac{k_BT}{\hbar\omega}\right)^3
g_{3}\left(\exp\left[\frac{\mu}{k_BT}\right],\exp\left[-\pi
\frac{r^2}{\lambda_T^2}\right]\right)\\
\label{g1volumeharm}
g^{(1)}({\bf r})&=&\frac{
g_{3}\left(\exp\left[\frac{\mu}{k_BT}\right],\exp\left[-\pi
\frac{r^2}{\lambda_T^2}\right]\right)}{g_{3}\left(\exp\left[\frac{\mu}{k_BT}
\right]\right)}.
\end{eqnarray}
The latter formula applies equally well to the case of an anisotropic harmonic
potential when the replacement 
$\omega=(\omega_x\omega_y\omega_z)^{1/3}$ is made.
As in the well known case of a spatially homogeneous system, the 
correlations decay on a length scale given by the thermal wavelength 
$\lambda_T$. However, the Bose function $g_3$ appropriate to a harmonic 
oscillator potential is closer to a linear function of its argument 
than the function $g_{3/2}$ characteristic of a homogeneous system.
The properties of a Bose-Einstein distribution are thus less visible in the
volume integrated correlation function of a harmonically trapped system
than in that of a homogeneous system, since the atoms 
away from the trap center experience a lower local chemical potential.
In view of this reduced tendency towards condensation for larger radii it may
be an acceptable approximation to assume a
Maxwell-Boltzmann occupation distribution of the trap levels, or equivalently 
to restrict the summation in Eq.\ (\ref{G1ho}) to its lowest order term. 
In this approximation we obtain the simple expressions
\begin{equation}
\label{gmb}
g^{(1)}({\bf r})=\exp\left(-\pi\frac{r^2}{\lambda_T^2}\right),\qquad
g^{(2)}({\bf r})=1+\exp\left(-2\pi\frac{r^2}{\lambda_T^2}\right).
\end{equation}
Numerical calculations confirm (cf.\ Fig.\ \ref{gidealfig}) the 
accuracy of these simple expressions even at temperatures very close to the 
critical temperature. 

\subsection{Temperatures below $T_c$}
\label{BelowTc}
In Sec.\ \ref{AboveTc} we have calculated the first and second order 
correlation functions for uncondensed systems. The presence of a Bose-Einstein
condensate, however, will lead to significant changes, especially in the
second order correlation function. 

In principle, the presence of a Bose-Einstein condensate has no 
effect on the previously discussed expression, Eq.\ (\ref{G1ideal}), for  
the first order correlation function $G^{(1)}$. However, due to the 
relatively long ranged coherence
of the condensate we have to perform the local density
approximation in a slightly different way. We therefore introduce 
a Wigner function, $W_T({\bf p},{\bf q})$, which describes only
the uncondensed atoms. 
The condensate, which corresponds to the ground state of the bare external
potential, is still considered to have a wave function $\psi({\bf r})$ that 
obeys the Schr\"odinger equation
\begin{equation}
\label{CondSE}
\left(-\frac{\hbar^2\nabla^2}{2m}+V({\bf r})-\mu\right)\,\psi({\bf r})=0,
\end{equation}
with $\mu=\hbar(\omega_x+\omega_y+\omega_z)/2$ corresponding to the zero
point energy of the oscillator potential.
The first order correlation function can then be decomposed into a condensed
and an uncondensed or thermal part
\begin{equation}
\label{G10}
G^{(1)}({\bf r},{\bf r}^\prime)=\psi^*({\bf r})\psi({\bf r}^\prime)+
G^{(1)}_T({\bf r},{\bf r}^\prime)
\end{equation}
with
\begin{equation}
\label{G1Wigner0}
G^{(1)}_T({\bf r},{\bf r}^\prime)=
\int\!d{\bf p}\, e^{-i{\bf p}\cdot
({\bf r}-{\bf r}^\prime)/\hbar}\,W_T({\bf p},\frac{{\bf r}+{\bf r}^\prime}{2}).
\end{equation}
Again, the thermal Wigner function $W_T$ describes the uncondensed part of the
full Wigner function.
The thermal correlation function $G^{(1)}_T$ is given by  
Eq.\ (\ref{G1lda}), since the momentum integration which led to Eq.\ 
(\ref{G1lda}) does not account for the existence of a condensate. However,
in contrast to Eq.\ (\ref{CondSE}), the chemical potential in $W_T$
is set equal to zero, ($\mu=0$), as it must be in the local density 
approximation.
The occupation of the condensate mode, furthermore, 
has to be determined from the normalization condition
\begin{equation}
\label{Nsum0}
N=\int\!d{\bf r}\,|\psi({\bf r})|^2+\int\!d{\bf r}\,G^{(1)}_T({\bf r},
{\bf r}).
\end{equation}
The limiting values of the normalized first order correlation function 
\begin{eqnarray}
g^{(1)}({\bf r},{\bf r})&=&1\\
\lim_{|{\bf r}-{\bf r}^\prime|\to\infty}
g^{(1)}({\bf r},{\bf r}^\prime)&=&\frac{\psi^*({\bf r})\,\psi({\bf r}^\prime)}
{\sqrt{G^{(1)}({\bf r},{\bf r})}
\,\sqrt{G^{(1)}({\bf r}^\prime,{\bf r}^\prime)}}
\end{eqnarray}
differ from those of an uncondensed gas (cf.\ Eq.\ (\ref{g1limits}))
by the possibility of off-diagonal long range order. The existence
of a condensate causes the correlation length of $g^{(1)}$ to be much larger 
than 
the thermal wavelength $\lambda_T$. In the case of a spatially homogeneous gas
the correlation extends throughout the entire system. The degree of first 
order coherence 
for large separations is then just given by the local condensate fraction.

The second order correlation function is affected by Bose-Einstein 
condensation in a more interesting way, as can be seen in Eq.\ 
(\ref{G2ideal}). The fluctuations of the number of atoms in the condensate
mode depend sensitively on the statistical ensemble that is assumed for the
calculation. The fluctuations characteristic of the grand canonical ensemble
greatly overestimate the condensate fluctuations in any experiment with
a fixed total atom number. Instead we assume canonical condensate
fluctuations which are of the order \cite{politzer96}
\begin{equation}
\langle\hat a_0^\dagger\hat a_0^\dagger\hat a_0\hat a_0\rangle =
\langle\hat a_0^\dagger\hat a_0\rangle\langle\hat a_0^\dagger\hat a_0\rangle
+O(N).
\end{equation}
In the limit of a large system we can neglect the last term since it is much 
smaller than the first term which is of the order $O(N^2)$. To calculate the
second order correlation function we refer back to Eq.\ (\ref{G2ideal}) and
note that the fluctuations of the uncondensed modes can be well approximated
by using the grand canonical ensemble. For the condensate mode, on the other
hand, we can write $\psi({\bf r})=\sqrt{\langle\hat a_0^\dagger\hat a_0\rangle}
\,u_0({\bf r})$, so that we have
\begin{eqnarray}
\label{G2belowTc}
G^{(2)}({\bf r},{\bf r}^\prime)&=& G^{(1)}({\bf r},{\bf r})\,
G^{(1)}({\bf r}^\prime,{\bf r}^\prime)
+ |G^{(1)}({\bf r},{\bf r}^\prime)|^2 - |\psi({\bf r})|^2\,|\psi({\bf r}^\prime
)|^2\\
\label{g2belowTc}
g^{(2)}({\bf r},{\bf r}^\prime)&=& 1 + |g^{(1)}({\bf r},{\bf r}^\prime)|^2
-\frac{|\psi({\bf r})|^2\,|\psi({\bf r}^\prime)|^2}{G^{(1)}({\bf r},{\bf r})\,
G^{(1)}({\bf r}^\prime,{\bf r}^\prime)}.
\end{eqnarray}
The limit for large relative separations is again given by
\begin{equation}
\lim_{|{\bf r}-{\bf r}^\prime|\to\infty}
g^{(2)}({\bf r},{\bf r}^\prime)=1.
\end{equation}
A certain measure \cite{dodd97} of second order coherence is 
\begin{equation}
\label{g2xxbelowTc}
g^{(2)}({\bf r},{\bf r})=2-\left(\frac{|\psi({\bf r})|^2}{G^{(1)}({\bf r},
{\bf r})}\right)^2.
\end{equation}
This function, which is determined by the ratio of the local condensate 
density to the total density must reduce to unity everywhere in order to 
describe perfect second order coherence. In general, it can only be close to
one near the center of the trap, where the condensate density is large. The
coherence properties of the thermal cloud outside the condensate, i.e.\ where
$\psi({\bf r})=0$, are not greatly different from those of an uncondensed gas.

\section{Strongly interacting, dilute gases}
\label{InteractingGas}
In Sec.\ \ref{IdealGas} we have studied the basic properties of the first and 
second order correlation functions of a trapped ideal gas. In this section
we discuss the way in which atomic interactions modify these results.
In order to deal with recent Bose condensation experiments 
\cite{anderson95,bradley95,davis95b} we focus 
on gases which are very dilute, but which contain atoms with a strong mutual 
repulsion. In this section we discuss only Bose condensed gases. The 
properties
of an uncondensed gas can easily be obtained as a specialisation of the
following treatment. 

\subsection{First order correlation function}
In the presence of atomic interactions, the condensate wave function obeys the
finite temperature Gross-Pitaevskii equation \cite{giorgini97} (cf.\ Eq.\ 
(\ref{CondSE}))
\begin{equation}
\label{GPE}
\left(-\frac{\hbar^2\nabla^2}{2m}+V({\bf r})+U\,[|\psi({\bf r})|^2
+2n_T({\bf r})]-\mu\right)\psi({\bf r})=0
\end{equation}
where the atomic interaction strength is measured by $U=4\pi\hbar^2a/m$,
with $a$ denoting the two-body scattering length. The local density of thermal
atoms $n_T({\bf r})$ is given by the still-to-be determined thermal 
part of the first order correlation function (cf.\ Eqs.\ (\ref{G10}) and
(\ref{G1Wigner0})),
\begin{equation}
\label{nTPopov}
n_T({\bf r})=G^{(1)}_T({\bf r},{\bf r})=\int d{\bf p}\,W_T({\bf p},{\bf r}).
\end{equation}
In the case of an interacting gas, the local density approximation has to be
applied not only to the distribution of thermal atoms but also to the 
condensate wave function in order to be consistent. Otherwise, the local
kinetic energy of the condensate wave function could be larger than the 
kinetic energy of the low lying elementary excitations. Details of this 
argument will be presented later. We have therefore to neglect the 
kinetic energy term in the Gross-Pitaevskii equation (\ref{GPE}), and that  
is appropriate if the kinetic energy is much smaller than the 
interaction energy. Under present experimental conditions \cite{giorgini97}
this condition is well fullfilled in the core of the condensate. 
However, the local density approximation can not be applied in a small 
region around the surface of the condensate, where the interaction energy
ceases to dominate the kinetic energy. The radial extent of this region 
is usually called the healing length. It has a fixed thickness and therefore a
vanishing relevance in the thermodynamic limit. This local 
density or 
Thomas-Fermi approximation has already been used frequently for the 
description of
Bose-Einstein condensation experiments. In this limit, the condensate 
wave function is given by
\begin{equation}
\label{PsiTF}
\psi({\bf r})=\sqrt{\frac{\mu-V({\bf r})-2\,U n_T({\bf r})}{U}}
\end{equation}
for positive arguments of the square root and by zero otherwise 
\cite{goldman81}.

Since the local density approximation fails close to the surface of the 
condensate we are only able to study correlation functions with both of their 
spatial arguments located within the condensate or outside of it. 
Inside the condensate, the first order correlation function is given by 
\begin{equation}
\label{G1i}
G^{(1)}({\bf r},{\bf r}^\prime)=\psi^*({\bf r})\psi({\bf r}^\prime)+
\int\!d{\bf p}\, e^{-i{\bf p}\cdot
({\bf r}-{\bf r}^\prime)/\hbar}\,W_T({\bf p},\frac{{\bf r}+{\bf r}^\prime}{2})
\end{equation}
with the condensate wave function defined by Eq.\ (\ref{PsiTF}).
For the calculation of the thermal Wigner function we have to take into account
that the low lying excited states of the interacting Bose gas are 
excitations of collective quasi-particle modes rather than single atoms. 
To be
more specific, a proper treatment of the atomic interactions in the presence
of a Bose-Einstein condensate starts with decomposing the atomic field 
operator into a condensate and a fluctuation part
\begin{equation}
\hat\Psi({\bf r})=\psi({\bf r})+\delta\hat\Psi({\bf r})
\end{equation}
with $\psi({\bf r})$ obeying the Gross-Pitaevskii equation (\ref{GPE}).
Due to this choice of $\psi({\bf r})$ only second order terms in 
the fluctuations $\delta\hat\Psi({\bf r})$ and higher contribute to the 
interaction Hamiltonian. Within
the Popov approximation third and fourth order terms in 
$\delta\hat\Psi({\bf r})$ are treated 
in an averaged way, neglecting anomalous expectation values like
$\langle\delta\hat\Psi({\bf r})\delta\hat\Psi({\bf r})\rangle$. This extends
the idea of the Bogoliubov approximation to finite temperatures since third 
and fourth
order fluctuation terms are completely neglected in the latter.
It is then possible to introduce quasi-particle operators
$\hat\alpha_j$ which diagonalize the approximate Hamilton operator of the 
interacting system to
\begin{equation}
\hat H=E_0+\sum_j\epsilon_j\,\hat\alpha_j^\dagger\hat\alpha_j.
\end{equation}
The quasi-particle operators $\hat\alpha_j$ obey Bose commutation relations.
In a spatially homogeneous system with volume $V$ they are related to the 
fluctuation operator $\delta\hat\Psi({\bf r})$ by \cite{fetter71}
\begin{equation}
\delta\hat\Psi({\bf r})=\frac{1}{\sqrt{V}}\sum_{\bf p}e^{i{\bf p}\cdot{\bf r}
/\hbar}
\left(u_{\bf p}\,\hat\alpha_{\bf p}
-v_{\bf p}\,\hat\alpha_{-\bf p}^\dagger\right).
\end{equation}
By virtue of the local density approximation both the coefficients 
$u({\bf p},{\bf r})$ 
and $v({\bf p},{\bf r})$ as well as the Hartree-Fock energy
\begin{eqnarray}
\epsilon_{\mbox{HF}}({\bf p},{\bf r})&=&\frac{p^2}{2m}+V({\bf r})
+2\,U\,[|\psi({\bf r})|^2
+n_T({\bf r})]-\mu\nonumber\\
\label{EHF}
&=&\frac{p^2}{2m}+U\,|\psi({\bf r})|^2
\end{eqnarray}
and the Popov energy
\begin{eqnarray}
\label{EPopov}
\epsilon({\bf p},{\bf r})&=&\sqrt{\left[\epsilon_{\mbox{HF}}
({\bf p},{\bf r})\right]^2-\left[U\,|\psi({\bf r})|^2\right]^2}\nonumber\\
&=&\sqrt{\left[\frac{p^2}{2m}+U\,|\psi({\bf r})|^2
\right]^2-\left[U\,|\psi({\bf r})|^2\right]^2}
\end{eqnarray}
become functions of the position ${\bf r}$. 
The local spectrum of elementary excitations $\epsilon({\bf p},{\bf r})$ 
coincides with the
well known Bogoliubov spectrum for a Bose condensed homogeneous gas 
close to zero temperature. The external potential and the interactions between
uncondensed atoms influence the excitation spectrum only indirectly 
through the local condensate density (cf.\ Eq.\ (\ref{PsiTF})).
If the kinetic energy of an elementary excitation with momentum 
${\bf p}$ is smaller than the interaction energy of the condensate atoms,
these excitations consist of phonon-like quasi-particles. For large kinetic 
energies, the
Popov energy $\epsilon({\bf p},{\bf r})$ coincides with the Hartree-Fock
energy $\epsilon_{\mbox{HF}}({\bf p},{\bf r})$ and the elementary excitations
then consist of real particles instead. 

Eq.\ (\ref{EPopov}) also explains
why it was necessary to extend the local density approximation to the 
condensate wave function. Had we not dropped the kinetic
energy of the condensate from the definition of the 
chemical potential in Eq.\ (\ref{GPE}), the Popov
energy would be given by
\begin{equation}
\epsilon({\bf p},{\bf r})=\sqrt{\left[\frac{p^2}{2m}+U\,|\psi({\bf r})|^2
+\frac{\hbar^2}{2m}\frac{\psi^*({\bf r})\nabla^2\psi({\bf r})}
{|\psi({\bf r})|^2}\right]^2-\left[U\,|\psi({\bf r})|^2\right]^2}.
\end{equation}
For a positive kinetic energy of the condensate wave function the
argument of the square root could become negative, leaving the Popov energy
undefined. One might try intuitively to remedy this problem, for example, by 
restricting the phase space available to the
quasi-particles. However, numerical comparisons of several such
schemes have yielded even less agreement with the results of an
exact Path Integral Monte Carlo calculations \cite{holzmann98b} 
than the complete neglect of the quasi-particle 
excitations that corresponds to the Hartree-Fock method.

Finally, we obtain the thermal Wigner function by assuming the 
quasi-particle states to be occupied by a Bose-Einstein distribution and by 
expressing the mode functions
$u({\bf p},{\bf r})$ and $v({\bf p},{\bf r})$ in terms of energies 
\cite{giorgini97}
\begin{eqnarray}
W_T({\bf p},{\bf r})&=&\frac{1}{(2\pi\hbar)^3}\left[u^2({\bf p},{\bf r})
+v^2({\bf p},{\bf r})\right]\frac{1}{\exp[
\epsilon({\bf p},{\bf r})/k_BT]-1}\nonumber\\
\label{WignerT0}
&=&\frac{1}{(2\pi\hbar)^3}\frac{\epsilon_{\mbox{HF}}
({\bf p},{\bf r})}{\epsilon({\bf p},{\bf r})}\frac{1}{\exp[
\epsilon({\bf p},{\bf r})/k_BT]-1}.
\end{eqnarray}

Outside the condensate, collective effects do not play any role in determining
the spectrum
of elementary excitations. The first order correlation function is then
given approximately by the formula (\ref{G1lda})
for an uncondensed ideal gas 
\begin{equation}
\label{G1ldai}
G^{(1)}({\bf r},{\bf r}^\prime)=\frac{1}{\lambda_T^3}g_{3/2}\left(\exp\left[
\frac{\mu[({\bf r}+{\bf r}^\prime)/2]}{k_BT}\right],\exp\left[-\pi
\frac{({\bf r}-{\bf r}^\prime)^2}{\lambda_T^2}\right]\right),
\end{equation}
in which the local chemical potential 
$\mu({\bf r})=\mu-V({\bf r})-\,2\,U n_T({\bf r})$ 
is reduced by the mean-field interaction between the uncondensed atoms. 
However, these interactions are usually very weak and the correlation function
is then essentially that discussed in Sec.\ \ref{IdealGas}.

\begin{figure}
\begin{center}
\leavevmode
\epsfxsize=0.5\textwidth
\epsffile{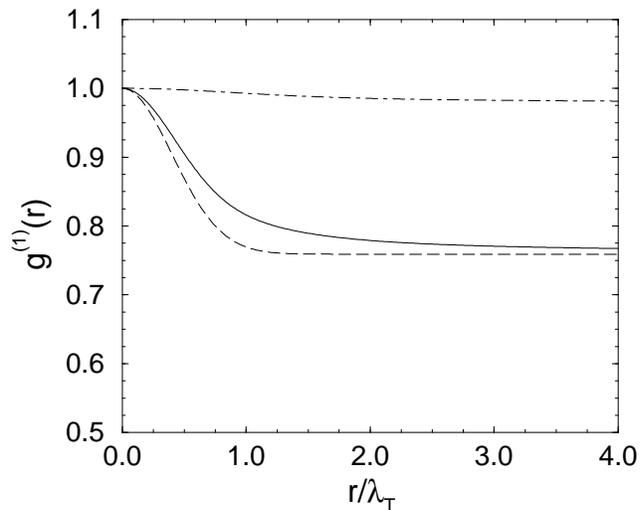}
\end{center}
\caption{\label{g1ifig}The solid line shows the volume integrated first order
correlation function for a harmonically trapped interacting gas at 
$T=0.5\,T_c^0$ where $T_c^0$ is the critical temperature of the ideal gas.
The scaling parameter is $\eta=0.31$. The correlation
function does not decay to zero due to the existence of a Bose-Einstein 
condensate. As in Fig.\ \ref{gidealfig}, the dashed line corresponds to a
Maxwell-Boltzmann distribution of the uncondensed atoms while the dot-dashed
line gives the unintegrated correlation function at the center of the trap.
The existence of quasi-particle excitations leads to a significantly slower
decay of the tail of the correlation function, compared to the case of
an ideal gas.}  
\end{figure}

A numerical plot of the volume integrated, normalized correlation function
$g^{(1)}({\bf r})$ for an interacting Bose condensed gas
is given in Fig.\ \ref{g1ifig}. The effective strength of 
the atomic interaction is best expressed in terms of the universal scaling 
parameter $\eta=\mu_{T=0}/k_BT_c^0$ \cite{giorgini97}, where $T_c^0$ is the 
critical temperature of the ideal gas. This scaling parameter assumes values
between $0.3$ and $0.4$ in most recent experiments. 
As in the case of a condensed ideal gas,
the correlation function decreases for relative separations much larger 
than the thermal wavelength, but smaller than the total size of the
condensate, to the
overall condensate fraction, i.e., the system posesses an off-diagonal long 
range order. However, for an interacting gas, the tail of the 
correlation function decreases much more slowly, as a result of the presence
of the long wavelength quasi-particle excitations.
The unintegrated correlation function $G^{(1)}({\bf r},{\bf r}^\prime)$ depends
strongly on its mean coordinate ${\bf R}=({\bf r}+{\bf r}^\prime)/2$
since the local condensate fraction varies significantly over the atomic 
sample. 

\subsection{Second order correlation function}
Atomic interactions affect the second order correlation function in a more
fundamental way than just by changing the spectrum of elementary excitations.
Atoms repel each other strongly at small separations. It is therefore 
impossible to find
two atoms at exactly the same location, i.e., $G^{(2)}({\bf r},{\bf r})=0$.
This effect is clearly not accounted for by Eqs.\ (\ref{G2belowTc}) and
(\ref{g2belowTc}). In order to study the influence of this hard core 
repulsion we briefly revisit the derivation of
the time-dependent version of the Gross-Pitaevskii equation (\ref{GPE}). 

The condensate dyad $\psi({\bf r})\psi^*({\bf r}^\prime)$ is
part of the reduced one-particle density matrix $\rho^{(1)}$, 
which contains the same information as the first order correlation function.
The dynamics of $\rho^{(1)}$ depends on the reduced two-particle density 
matrix 
$\rho^{(2)}$ through the Schr\"odinger equation
\begin{equation}
\label{BBGKY}
\frac{\partial}{\partial t}\rho^{(1)}=-\frac{i}{\hbar}\big[H_{0,1},\rho^{(1)}
\big]
-\frac{i}{\hbar}\mbox{tr}_2\left\{\big[H_I,\rho^{(2)}\big]\right\}
\end{equation}
with the one-particle Hamiltonian $H_{0,1}$ and the true atomic interaction
potential $U({\bf r}_1-{\bf r}_2)$ in the coordinate representation
\begin{equation}
H_{0,i} = -\frac{\hbar^2\nabla^2_i}{2m}+V({\bf r}_i),\qquad
H_I=U({\bf r}_1-{\bf r}_2).
\end{equation}
The dynamics of the second order correlation function is similary linked to
the third-order correlation function and so on, the entire set of 
$N$ coupled equations constituting the so-called BBGKY hierarchy. 
If it is possible, however, to express the reduced two-particle density matrix
$\rho^{(2)}$ in terms of the reduced one-particle density matrix $\rho^{(1)}$, 
the hierarchy can be truncated by turning
Eq.\ (\ref{BBGKY}) into a closed nonlinear equation. 

In most treatments, the time-dependent analog of the Gross-Pitaevskii 
equation (\ref{GPE}) is obtained by replacing the true atomic 
interaction potential by the point interaction $\tilde H_I = 
U\,\delta({\bf r}_1-{\bf r}_2)$ and by factorizing the reduced two-particle
density matrix analogously to Eq.\ (\ref{G2belowTc}). 
Even though such an application of the first order Born approximation to a 
suitably chosen pseudopotential properly reproduces the Gross-Pitaevskii 
equation, its implications for quantities other than
the reduced one-particle density matrix require more careful examination.
For example, the above treatment implies that the second order 
correlation function  
\begin{equation}
\label{G20}
G^{(2)}_0({\bf r},{\bf r}^\prime)=G^{(1)}({\bf r},{\bf r})\,
G^{(1)}({\bf r}^\prime,{\bf r}^\prime)
+ |G^{(1)}({\bf r},{\bf r}^\prime)|^2 - |\psi({\bf r})|^2\,|\psi({\bf r}^\prime
)|^2
\end{equation}
corresponds to an ideal gas of quasi-particles with 
$G^{(1)}({\bf r},{\bf r}^\prime)$ given by Eq.\ (\ref{G1i}). 
Such a conclusion, however, incorrectly ignores the strong 
repulsion between atoms at small distances. 

We therefore outline a more detailed derivation of the 
Gross-Pitaevskii equation \cite{naraschewski97b,balescu75}
which also describes the two-particle density matrix
appropriately. For this purpose we
decompose the two-particle density matrix, $\rho^{(2)}=\rho^{(2)}_0+\delta
\rho^{(2)}$, into a part $\rho^{(2)}_0$ representing
products of the reduced single-particle density matrix (cf.\ Eq.\ (\ref{G20})) 
and a part
$\delta\rho^{(2)}$, which corrects for the true two-particle correlations. 
Rather than neglecting the correlation part $\delta\rho^{(2)}$ altogether, 
we turn Eq.\ (\ref{BBGKY}) into a closed equation by solving 
the dynamics of $\delta\rho^{(2)}$ analytically and by reexpressing 
the correlations, $\delta\rho^{(2)}$, in terms of
$\rho^{(2)}_0$. Of course, certain approximations have to be made in order to
achieve a convenient analytical expression for $\delta\rho^{(2)}$.

The equation of motion of $\delta\rho^{(2)}$ is simplified by 
the assumption of an atom density low enough to permit neglect of 
three-body and higher order collisions. The two-particle correlations are
therefore the result of two-particle scattering only. 
In the context of many-particle physics this approach is often called a ladder
approximation since the two-body collision is calculated up to arbitrary 
orders in the true interaction Hamiltonian $H_I$. 
It is worth noting that a literal
point-like potential does not lead to any two-particle scattering, and that
is consistent with the assumption $\rho^{(2)}=\rho^{(2)}_0$, 
but it is not at all realistic.

If we describe the two-particle scattering by the M{\o}ller operator
\begin{equation}
\Omega = \lim_{t\to\infty}\exp[-i(H_{0,1}+H_{0,2}+H_I)t/\hbar]
\exp[i (H_{0,1}+H_{0,2})t/\hbar],
\end{equation}
it can be shown that the properly approximated two-particle
density matrix is given by $\rho^{(2)}=\Omega\,\rho^{(2)}_0\,\Omega^\dagger$.
Thus the equation of motion for $\rho^{(1)}$ reads \cite{balescu75}
\begin{equation}
\frac{\partial}{\partial t}\rho^{(1)}=-\frac{i}{\hbar}\big[H_{0,1},\rho^{(1)}
\big]-\frac{i}{\hbar}\mbox{tr}_2\left\{\big[H_I,\Omega\,\rho^{(2)}_0\,
\Omega^\dagger\big]\right\}.
\end{equation}
The M{\o}ller operator $\Omega$ 
is related to the two-body $T$-matrix by the relation $T=H_I\,\Omega$.
The matrix elements of the two-body $T$-matrix in momentum representation are
identical to the scattering amplitude.
It is therefore the $T$-matrix rather than the atomic interaction potential
$H_I$ which should be approximated by
a delta function $T=U\,\delta({\bf r}_1-{\bf r}_2)$. To lowest order in the 
scattering length, this equation of motion can be further reduced to
\begin{equation}
\label{G1kinetic}
\frac{\partial}{\partial t}\rho^{(1)}
\approx-\frac{i}{\hbar}\big[H_{0,1},\rho^{(1)}\big]
-\frac{i}{\hbar}\mbox{tr}_2\left\{\big[T,\rho^{(2)}_0\big]\right\}
\end{equation}
which coincides with the previously outlined mean-field approximation. The 
neglected terms, which are quadratic in the scattering length, describe small
thermal energy shifts and the dissipative effects of collisions such as 
thermalization \cite{naraschewski97b}. 

We conclude from the above derivation of the Gross-Pitaevskii equation 
that the second order correlation function
that properly accounts for hard core repulsion is given by
\begin{equation}
\label{G2scatt}
\hat G^{(2)}=\Omega^\dagger\,\hat G^{(2)}_0\,\Omega.
\end{equation}
The correlation function is here defined as the operator 
$\hat G^{(2)}=\rho^{(2)\,\dagger}$ with the understanding that its coordinate
representation corresponds to $G^{(2)}({\bf r},{\bf r}^\prime)$.

For reasons of simplicity we calculate $G^{(2)}$ first for an uncondensed gas.
The scattering involves only the wave function for the relative coordinate
${\bf r}={\bf r}_1-{\bf r}_2$ of the two 
atoms. If the incoming states are plane waves
the effect of the M{\o}ller operator is given by \cite{sakurai94}
\begin{equation}
\Omega\,e^{i({\bf p_1}\cdot{\bf r}_1+{\bf p_2}\cdot{\bf r}_2)/\hbar}=
e^{i{\bf P}\cdot{\bf R}/\hbar}\left[
e^{i{\bf p}\cdot{\bf r}/\hbar}-a\frac{e^{ipr/\hbar}}{r}\right]\qquad(r\gg a)
\end{equation}
with the center-of-mass coordinate ${\bf R}=({\bf r_1}+{\bf r}_2)/2$, the
total momentum ${\bf P}={\bf p}_1+{\bf p}_2$, and the relative momentum 
${\bf p}=({\bf p}_1-{\bf p}_2)/2$.
Here, we have assumed that all relative momenta are small enough to permit
taking only
the momentum independent part, $-a$, of the scattering amplitude into account.
With this approximation we have neglected terms of order
$pa/\hbar$ which is justified if $a\ll\lambda_T$. In a harmonic trap the 
ratio $a/\lambda_T$ is given by
\begin{equation}
\frac{a}{\lambda_{T}}=\frac{4}{15}\frac{1}{\sqrt{\pi}}\frac{1}{\zeta(3)}
\eta^{5/2}\left(\frac{T}{T_c}\right)^{1/2}.
\end{equation}
It is mainly determined by the scaling
parameter $\eta=\mu_{T=0}/k_BT_c$ \cite{giorgini97} which, as we have noted,
assumes values between $0.3$ and $0.4$ in most recent experiments.
As a consequence, we can generally assume a ratio of 
$a/\lambda_T\approx 0.01$ at the critical temperature.

In agreement with the local density approximation,
we evaluate $G^{(2)}$ for a spatially homogeneous gas with volume $V$.
Denoting the relative coordinate once more as ${\bf r}$ and the relative 
momentum as ${\bf p}$, we obtain
for an uncondensed system (cf.\ Eq.\ (\ref{G2aboveTc}))
\begin{eqnarray}
G^{(2)}({\bf r}_1,{\bf r}_2)&=&\frac{1}{V^2}\sum_{{\bf p}_1,{\bf p}_2}
\left(\Omega\,e^{i({\bf p_1}\cdot{\bf r}_1+{\bf p_2}\cdot{\bf r}_2)/\hbar}
\right)^\dagger\left(\Omega\,e^{i({\bf p_1}\cdot{\bf r}_1+{\bf p_2}\cdot
{\bf r}_2)/\hbar}\right)\langle n_{{\bf p}_1}\rangle\langle n_{{\bf p}_2}
\rangle\nonumber\\
&+&\frac{1}{V^2}\sum_{{\bf p}_1,{\bf p}_2}
\left(\Omega\,e^{i({\bf p_1}\cdot{\bf r}_1+{\bf p_2}\cdot{\bf r}_2)/\hbar}
\right)^\dagger\left(\Omega\,e^{i({\bf p_2}\cdot{\bf r}_1+{\bf p_1}\cdot
{\bf r}_2)/\hbar}\right)\langle n_{{\bf p}_1}\rangle\langle n_{{\bf p}_2}
\rangle\nonumber\\
&=&\frac{1}{V^2}\sum_{{\bf p}_1,{\bf p}_2}
\left(e^{i{\bf p}\cdot{\bf r}/\hbar}-a\frac{e^{ipr/\hbar}}{r}\right)^\dagger
\left(e^{i{\bf p}\cdot{\bf r}/\hbar}-a\frac{e^{ipr/\hbar}}{r}\right)\langle 
n_{{\bf p}_1}\rangle\langle n_{{\bf p}_2}\rangle\nonumber\\
&+&\frac{1}{V^2}\sum_{{\bf p}_1,{\bf p}_2}
\left(e^{i{\bf p}\cdot{\bf r}/\hbar}-a\frac{e^{ipr/\hbar}}{r}\right)^\dagger
\left(e^{-i{\bf p}\cdot{\bf r}/\hbar}-a\frac{e^{ipr/\hbar}}{r}\right)
\langle n_{{\bf p}_1}\rangle\langle n_{{\bf p}_2}
\rangle.
\end{eqnarray}
After multiplying this expression out term by term and making use of the 
spatial isotropy
of the homogeneous gas we can write the second order correlation function as
\begin{eqnarray}
G^{(2)}({\bf r},{\bf r}^\prime)&=&G^{(1)}({\bf r},{\bf r})\,
G^{(1)}({\bf r}^\prime,{\bf r}^\prime)\left(1+\frac{2a^2}{|{\bf r}-{\bf r}
^\prime|^2}\right)+|G^{(1)}({\bf r},{\bf r}^\prime)|^2
\left(1-\frac{4a}{|{\bf r}-{\bf r}^\prime|}\right)\nonumber\\
&=&\left[G^{(1)}({\bf r},{\bf r})\,G^{(1)}({\bf r}^\prime,{\bf r}^\prime)
+|G^{(1)}({\bf r},{\bf r}^\prime)|^2\right]\left(1-\frac{a}
{|{\bf r}-{\bf r}^\prime|}\right)^2\nonumber\\
\label{G2i}
&&+\left[G^{(1)}({\bf r},{\bf r})\,G^{(1)}({\bf r}^\prime,{\bf r}^\prime)
-|G^{(1)}({\bf r},{\bf r}^\prime)|^2\right]\left[\left(1+\frac{a}
{|{\bf r}-{\bf r}^\prime|}\right)^2-1\right],
\end{eqnarray}
an expression
which should also be valid for a sufficiently slowly varying inhomogeneous 
system. The first line of Eq.\ (\ref{G2i}) multiplies the correlation function
of an ideal gas by the squared 
relative wave function in the limit of zero momentum scattering. 
The additional term on the second line accounts for the fact that the 
atoms have finite relative momenta with a distribution implicitly
characterized by $G^{(1)}({\bf r},{\bf r}^\prime)$. This correction term can
only be neglected for very small $a/\lambda_T$ ratios. Furthermore, only
linear terms in $a/|{\bf r}-{\bf r}^\prime|$ are relevant for distances 
$|{\bf r}-{\bf r}^\prime|\gg a$.

The extension of the previous calculation to Bose condensed gases is 
straightforward
though it requires a little more algebra due to the existence
of quasi-particle excitations. However, most of the required modifications are
absorbed into the definition of the first order correlation function 
$G^{(1)}$. In addition one has to
take into account the fact that the second order correlation function of a 
condensed ideal gas
is given by Eq.\ (\ref{G2belowTc}) rather than by Eq.\ (\ref{G2aboveTc}). With
these modifications the second order correlation function of a condensed
interacting gas is given by
\begin{eqnarray}
\label{G2i0}
G^{(2)}({\bf r},{\bf r}^\prime)
&=&G^{(2)}_0({\bf r},{\bf r}^\prime)\,
\left(1-\frac{a}{|{\bf r}-{\bf r}^\prime|}\right)^2\\
&&\qquad+\left[G^{(1)}({\bf r},{\bf r})\,G^{(1)}({\bf r}^\prime,{\bf r}^\prime)
-|G^{(1)}({\bf r},{\bf r}^\prime)|^2\right]\left[\left(1+\frac{a}
{|{\bf r}-{\bf r}^\prime|}\right)^2-1\right]\nonumber\\
g^{(2)}({\bf r},{\bf r}^\prime)
&=&g^{(2)}_0({\bf r},{\bf r}^\prime)\,
\left(1-\frac{a}{|{\bf r}-{\bf r}^\prime|}\right)^2
+\left[1-|g^{(1)}({\bf r},{\bf r}^\prime)|^2\right]\left[\left(1+\frac{a}
{|{\bf r}-{\bf r}^\prime|}\right)^2-1\right].
\end{eqnarray}

\begin{figure}
\begin{center}
\leavevmode
\epsfxsize=0.5\textwidth
\epsffile{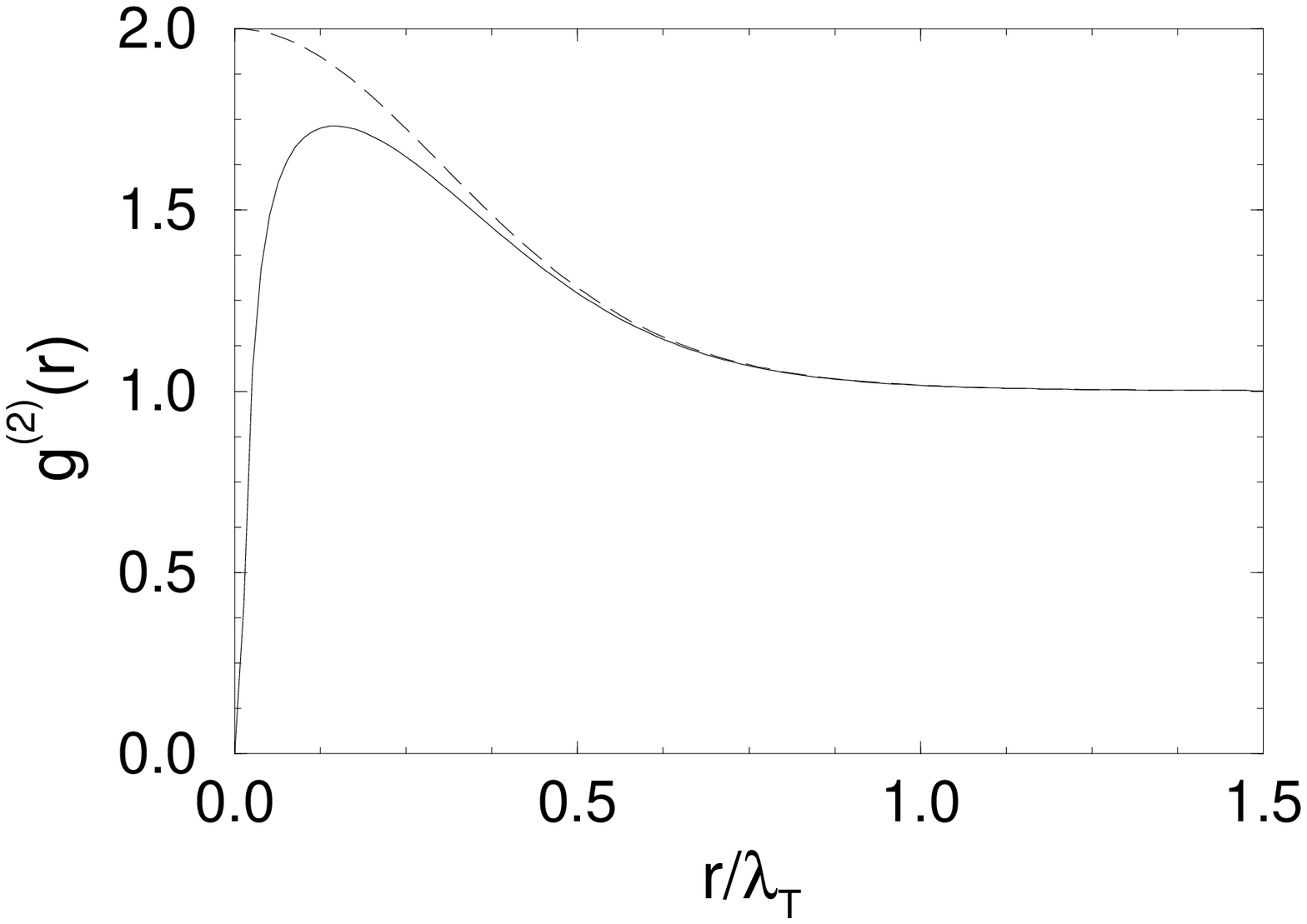}
\end{center}
\caption{\label{g2i10fig}The volume integrated second order correlation
function of an interacting gas is given by the solid line. The temperature
corresponds to the critical temperature $T_c^0$ of an ideal gas. With a 
scaling parameter of $\eta=0.31$, the scattering length $a$ is much smaller
than the thermal wavelength,
$a=0.007\,\lambda_T$. Nevertheless, the hard core repulsion leads to
noticeable reduction of the correlation function over a broad range of
relative distances. The dashed line shows as a comparison the volume 
integrated fictitious
correlation function $g^{(2)}_0({\bf r})$ which omits the effect of the hard 
core repulsion.}  
\end{figure}

\begin{figure}
\begin{center}
\leavevmode
\epsfxsize=0.5\textwidth
\epsffile{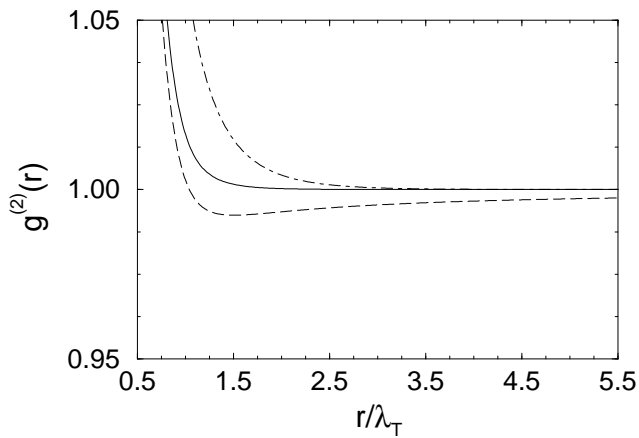}
\end{center}
\caption{\label{g2i10afig}This plot compares the long range behavior of the 
volume integrated second
order correlation function of Fig.\ \ref{g2i10fig} (solid line) with its main
constituent $g_0^{(2)}(r)(1-a/r)^2$ (dashed line). The latter approximates
the relative wave function by its zero temperature limit. 
The difference is small, and becomes even smaller at lower temperatures. 
The dot-dashed line, by contrast, displays the unintegrated correlation 
function at the center of the trap.}
\end{figure}

\begin{figure}
\begin{center}
\leavevmode
\epsfxsize=0.5\textwidth
\epsffile{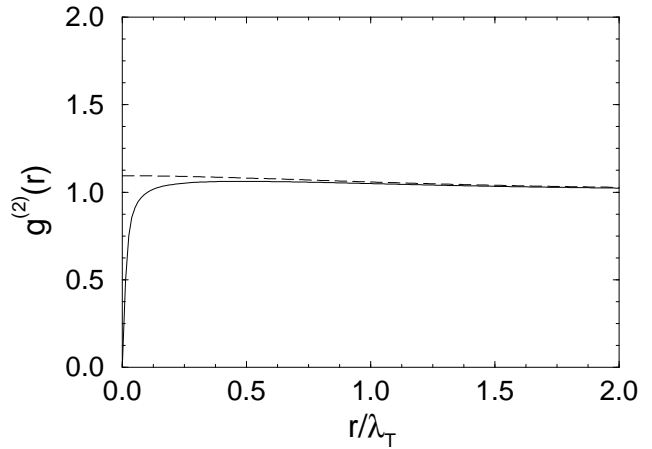}
\end{center}
\caption{\label{g2i05fig}The volume integrated second order correlation
function is here plotted for a temperature $T=0.5\,T_c^0$. The dashed line
shows $g^{(2)}_0({\bf r})$ which omits the hard core repulsion.
Due to the presence of a considerable condensate fraction, a large degree
of second order coherence is attained. The existence of quasi-particle
excitations leads to a remarkably slow decrease of the correlation function.} 
\end{figure}

In Fig.\ \ref{g2i10fig} we compare the volume integrated normalized 
correlation functions 
$g^{(2)}$ and $g_0^{(2)}$ for an uncondensed system at the critical 
temperature of an ideal gas $T_c^0$. The interaction 
strength is given by the scaling parameter $\eta=0.31$. Even though the
scattering length is much smaller than the thermal 
wavelength, with $a=0.007\lambda_T$, the hard core repulsion significantly 
reduces the amount of
bosonic attraction between the atoms. It becomes apparent from Fig.\ 
\ref{g2i10afig} that $g^{(2)}({\bf r})$ is dominated by  
$g_0^{(2)}({\bf r})(1-a/r)^2$ corresponding to the first line of Eq.\ 
(\ref{G2i}). The additional terms contribute a $1\%$ correction.
They mainly suppress the residual effect of the hard core repulsion for 
distances of the order of the thermal wavelength and larger. Fig.\ 
\ref{g2i10afig} also illustrates the diluting effect of the volume integration
on the range of the bosonic attraction.
For comparison, we plot in Fig.\ \ref{g2i05fig} the correlation function 
$g^{(2)}$ for a Bose condensed gas at $T=0.5\,T_0$. Its maximum value is 
strongly reduced since condensate atoms do not exhibit any tendency to clump
together due to wave function symmetrization. On the other hand, the
existence of quasi-particle excitations leads to a relatively long tail for 
the correlation function. For this temperature, corrections to the term 
$g_0^{(2)}({\bf r})(1-a/r)^2$ can be safely neglected,  
due to a larger thermal wavelength, i.e., 
$a/\lambda_T=0.005$, and the long tail of the correlation function.
 
\section{Coherence properties}
\label{CoherenceProp}
We summarize in this section the coherence properties of trapped Bose gases
based on our results for the first and second order correlation 
functions. First order coherence describes the ability of atoms with an initial
separation $\bf r$ to form an interference pattern, when brought together.
For an uncondensed gas, the visibility of such an interference pattern drops
to zero on a length scale of $\bf r$ measured by the thermal wavelength, 
which is roughly of the order of $0.5\mu$m in present experiments. 
In a Bose condensed gas, however, the visibility decreases on the same length
scale to a plateau given by the total fraction of condensed 
atoms. This off-diagonal long range order allowed the experimental 
observation of an interference pattern \cite{andrews97} which extended over 
the whole cloud of condensate atoms. For spatial separations larger than the
the size of the condensate, the visibility drops to zero as well, 
because long range order can not extend beyond the system size.

Second order coherence implies instead the absence of any density
correlations other than the trivial autocorrelation pointed out in Eq.\
(\ref{ncorr}). In a dilute Bose gas, we can identify two sources of density 
correlations which limit the achievable degree of second order coherence.
An inevitable anticorrelation between atoms originates from their strong mutual
repulsion. The length scale of this repulsion is given by the 
scattering length $a$ which is of the order of a few nanometers and therefore
much smaller than any other relevant length scale. However, the relative 
wave function of two atoms increases as $a/r$, which is to say very slowly. 
Anticorrelation effects are 
therefore still visible on the length scale of the thermal wavelength, which
is about 200 times larger (at $T=0.5\,T_c^0$) than the scattering length. 
The other limitation to second order coherence is due to the tendency of 
bosonic atoms to cluster together. The range of this effective bosonic 
attraction is again limited by the thermal wavelength. However, its maximal 
amount is a function of temperature too, since the condensate atoms do not 
exhibit such a clustering effect. Because of the interplay between hard core
repulsion and bosonic clustering, the maximum of the second order correlation
function depends on both the temperature and the scattering length. 

\begin{figure}
\begin{center}
\leavevmode
\epsfxsize=0.5\textwidth
\epsffile{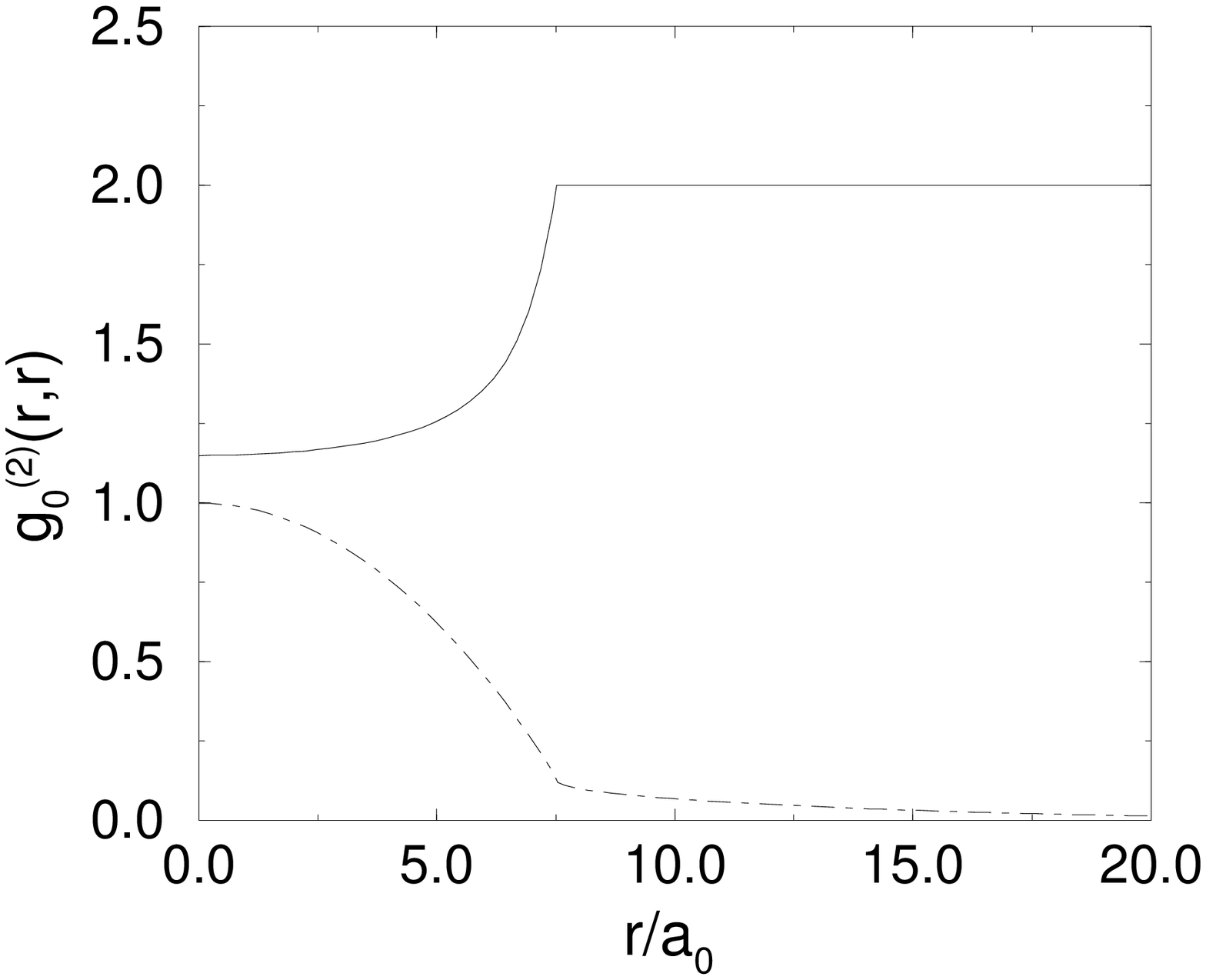}
\end{center}
\caption{\label{g2xxifig} The normalized second order correlation function
$g^{(2)}_0({\bf r},{\bf r})$ of Eq.\ (\ref{g2xxbelowTc})
is a measure of local second order coherence. It also describes the ratio
of the total interaction energy, including quantum statistical exchange energy,
to the classical Hartree mean-field energy. We have plotted 
$g^{(2)}_0({\bf r},{\bf r})$ (solid line) for a harmonically trapped gas with 
interaction strength $\eta=0.31$ and temperature $T=0.8\,T_c^0$. 
For comparison, the dot-dashed curve shows the total density distribution 
in arbitrary units.}  
\end{figure}

It is nevertheless useful to characterize the local second order coherence by 
the maximum of the fictitious correlation function $g_0^{(2)}({\bf r},
{\bf r})$ which does not
account for hard core repulsion.  We illustrate in Fig.\ \ref{g2xxifig} how 
this local measure of second order coherence, which is a function of the
local condensate density, varies with the radial coordinate $r$ in 
a spherically symmetric system (cf.\ \cite{dodd97}). Analogously, the 
maximum $g_0^{(2)}(0)$ of the volume integrated  
correlation function offers a certain global measure of the second order 
coherence
of the system. Fortunately, the quantity $g_0^{(2)}(0)$ is 
experimentally accessible due to its relation to the total atomic interaction
energy. According to Eq.\ (\ref{G1kinetic}) the latter energy can be 
approximated by
\begin{equation}
\langle H_I\rangle/2 = \mbox{tr}\left\{H_I\,G^{(2)}\right\}/2
\approx\mbox{tr}\left\{T\,G^{(2)}_0\right\}/2 = \frac{U}{2} 
\int\!d{\bf r}\,G^{(2)}_0({\bf r},{\bf r})=\frac{U}{2}\,G_0^{(2)}(0).
\end{equation}
The overlap integral between the true atomic interaction potential and the
relative wave function between the atoms is therefore approximately 
proportional to the maximum value of the otherwise fictitious volume integrated
correlation function $G_0^{(2)}({\bf r})$. 
 
In a recent experiment \cite{ketterle97} the interaction energy and thereby 
$g_0^{(2)}(0)$ were determined
by measuring the release energy of the atoms after switching off the magnetic
trap. This experiment provided some evidence for a large degree of 
second order coherence of a trapped Bose-Einstein condensate at zero
temperature. At finite temperature, however, a 
determination of the interaction energy 
from the release energy would require
that the initial kinetic energy of the uncondensed atoms be measured 
independently.
In a similar experiment the third order analog of $g_0^{(2)}(0)$ has been
determined by measuring three-body scattering rates \cite{burt97}. 
The implications of these experiments for second and third order coherence has
been discussed before \cite{ketterle97,burt97,dodd97}. However, it was wrongly
concluded that these experiments measure the true
second order correlation function $G^{(2)}$ or its third order analog. 
The significant impact of the hard core repulsion on second and
third order coherence has apparently not yet been fully appreciated.

\begin{figure}
\begin{center}
\leavevmode
\epsfxsize=0.5\textwidth
\epsffile{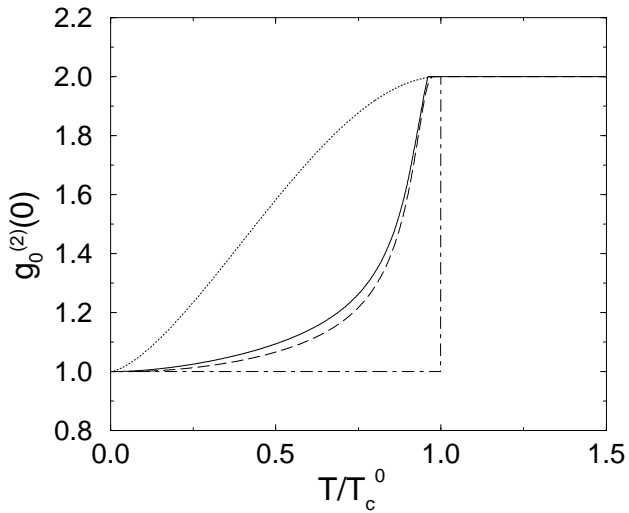}
\end{center}
\caption{\label{g200ifig} The peak value $g^{(2)}_0(0)$ of the volume 
integrated second order correlation function, excluding the hard core repulsion
between the atoms, is a good measure of the second order coherence
of the system. The solid line shows the influence of the temperature on the 
second order coherence for a harmonically trapped interacting gas in the
Popov approximation with scaling parameter $\eta = 0.31$. A corresponding 
calculation using the Hartree-Fock approximation (dashed line) shows that
quasi-particle excitations have a small but noticeable effect.
For comparison, the dotted line describes a spatially homogeneous ideal gas
while the dot-dashed line corresponds to a harmonically trapped ideal gas.  
}  
\end{figure}

Finally, we discuss the temperature dependence of $g_0^{(2)}(0)$ in 
Fig.\ \ref{g200ifig}. The numerical plot shows a rapid but continuous decrease
of $g_0^{(2)}(0)$ as the temperature falls below the critical temperature.
A comparison with a calculation in the Hartree-Fock limit shows that the 
existence of quasi-particle excitations has a small but noticeable effect on
second order coherence. We contrast these results with the second order
coherence of both a spatially homogeneous and a harmonically trapped ideal 
Bose gas. In a spatially homogeneous medium one finds
\begin{equation}
\label{g20hom}
g^{(2)}(0)=2-\left(\frac{N_0}{N}\right)^2=2-\left[1-\left(\frac{T}{T_c}
\right)^{3/2}\right]^2,
\end{equation}
which is almost a linear function of temperature (cf.\ Fig.\ \ref{g200ifig}).
In contrast the volume integrated second order correlation 
function of a harmonically trapped ideal Bose gas 
\begin{eqnarray}
g^{(2)}(0)&=&2-\frac{\int\!\!d{\bf r}\,|\psi({\bf r})|^4}
{\int\!\!d{\bf r}\,G^{(1)}({\bf r},{\bf r})^2},\nonumber\\
&=&1+O(1/\sqrt{N}),\qquad T\le T_c
\end{eqnarray}
experiences a sharp transition from the value two above the critical 
temperature to the value one below it.
The reason for this discontinuous transition is that the peak density of
a harmonically trapped noninteracting Bose gas becomes infinite as the
thermodynamic limit is reached while the peak density of the thermal cloud
retains a finite value. Of course, the presence of 
atomic interactions prevents any such unbounded growth of the condensate 
density for a real gas.
Nevertheless, the spatial inhomogeneity accelerates the transition to
a second order coherent ensemble as the temperature is decreased.

\section{Conclusion}
\label{Conclusion}
We have extended the concept of optical coherence, defined in terms
of correlation functions, to dilute trapped Bose gases.
Except for a small region around the surface of a 
Bose-Einstein condensate, these correlation functions can be calculated by 
using a local density approximation. Even though the system
behaves locally as if it were spatially homogeneous, the spatial dependence of
the effective chemical potential adds some interesting new physical features.
The influence of both the Bose-Einstein distribution of uncondensed 
atoms and of quasi-particle like
excitations is strongly reduced when a volume average over the mean coordinate
of the correlation functions is performed. Nevertheless, consequences of 
quasi-particle excitations are still noticeable in the tails of the correlation
functions. The density correlation is dominated by a tendency of bosonic
atoms to cluster together and, on the other hand, the existence of a 
hard core repulsion between these atoms. The quantum 
statistical attraction and the effect of coulomb repulsion are found to be of
comparable range for the conditions of most recent experiments.
It is in this sense incorrect to neglect
the hard core repulsion in the discussion of density correlations. 

For the discussion of second order coherence, however, it is convenient to 
introduce a
fictitious correlation function which does omit the effect of hard core
repulsion. This correlation function is experimentally accessible due to its 
close relation to the atomic interaction energy. By introducing this fictitious
correlation function, we are able to explain the results of recent 
experiments on second and third order coherence.
Finally, we have observed that a harmonically
trapped gas achieves second order coherence much more rapidly as the 
temperature is lowered below the critical temperature than does its spatially 
homogeneous counterpart.

\acknowledgments
It is a pleasure to thank M. Holzmann for stimulating discussions and the
opportunity to compare our calculations with the unpublished results of a
numerical Path Integral Monte Carlo calculation \cite{holzmann98a}.
M.N. acknowledges financial support by the Deutsche Forschungsgemeinschaft.  


\narrowtext
\twocolumn

\end{document}